\documentclass[aps,pra,twocolumn,superscriptaddress,showpacs]{revtex4-1}
\usepackage{graphicx}
\usepackage{dcolumn}
\usepackage{bm}
\usepackage{color}
\usepackage{amssymb} 
\usepackage{amsmath}
\usepackage{verbatim}
\usepackage{float}
\usepackage{subfig}
\usepackage{lmodern}
\usepackage[utf8]{inputenc}
\usepackage[T1]{fontenc}

\usepackage{ragged2e}
\DeclareCaptionJustification{justified}{\justifying}
\renewcommand\Re{\operatorname{Re}}
\renewcommand\Im{\operatorname{Im}}
\newcommand{\la}{\langle}
\newcommand{\ra}{\rangle}
\newcommand{\rt}{\right}
\newcommand{\lf}{\left}
\newcommand{\Tr}{\operatorname{Tr}}

\renewcommand{\k}{\mathbf k}
\newcommand{\s}{\text{s}}
\renewcommand{\i}{\text{in}}

\newcommand{\el}{\text{el}}

\begin{document}
\title{Imaging electron dynamics with time- and angle-resolved photoelectron spectroscopy}
\author{Daria Popova-Gorelova}
\email[]{daria.gorelova@desy.de}
\affiliation{Center for Free-Electron Laser Science, DESY, Notkestrasse 85, D-22607 Hamburg, Germany}
\affiliation{The Hamburg Centre for Ultrafast Imaging, University of Hamburg, Luruper Chaussee 149, D-22761 Hamburg, Germany}
\author{Jochen K\"upper}
\affiliation{Center for Free-Electron Laser Science, DESY, Notkestrasse 85, D-22607 Hamburg, Germany}
\affiliation{The Hamburg Centre for Ultrafast Imaging, University of Hamburg, Luruper Chaussee 149, D-22761 Hamburg, Germany}
\affiliation{Department of Physics, University of Hamburg, Luruper Chaussee 149, D-22761 Hamburg, Germany}
\author{Robin Santra}
\email[]{robin.santra@cfel.de}
\affiliation{Center for Free-Electron Laser Science, DESY, Notkestrasse 85, D-22607 Hamburg, Germany}
\affiliation{The Hamburg Centre for Ultrafast Imaging, University of Hamburg, Luruper Chaussee 149, D-22761 Hamburg, Germany}
\affiliation{Department of Physics, University of Hamburg, Jungiusstrasse 9, D-20355 Hamburg, Germany} 
\date{\today}

\begin{abstract}

We theoretically study how time- and angle-resolved photoemission spectroscopy can be applied for imaging coherent electron dynamics in molecules. We consider a process in which a pump pulse triggers coherent electronic dynamics in a molecule by creating a valence electron hole. An ultrashort extreme ultraviolet (XUV) probe pulse creates a second electron hole in the molecule. Information about the electron dynamics is accessed by analyzing angular distributions of photoemission probabilities at a fixed photoelectron energy. We demonstrate that a rigorous theoretical analysis, which takes into account the indistinguishability of transitions induced by the ultrashort, broadband probe pulse and electron hole correlation effects, is necessary for the interpretation of time- and angle-resolved photoelectron spectra. We show how a Fourier analysis of time- and angle-resolved photoelectron spectra from a molecule can be applied to follow its electron dynamics by considering photoelectron distributions from an indole molecular cation with coherent electron dynamics.

\end{abstract}
\maketitle

\section{Introduction}

Electron dynamics in valence shells of atoms and molecules determine various chemical and physical transformations in solids and molecules. Therefore, real-time imaging of electron dynamics is one of the most important goals for modern ultrafast science \cite{KrauszRMP09, SmirnovaNature09, GoulielmakisNature10, SansoneNature10, HaesslerNature10, TzallasNature11, HockettNature11,  BlagaNature12, PullenNature15}. In this article, we investigate the opportunities for applying ultrashort extreme ultraviolet (XUV) probe pulses inducing single-photon ionization for imaging coherent electron dynamics in molecules by means of time- and angle-resolved photoelectron spectroscopy (TRARPES), {\it i.e.}, time- and energy-resolved molecular-frame photoelectron angular distributions.

An advantage of photoelectron imaging over x-ray imaging is that it allows achieving \r Angstrom spatial resolution with light pulses of a much lower photon energy. And its advantage over ultrafast electron diffraction is that much higher temporal resolution can be obtained using light pulses than using electron pulses \cite{KrauszRMP09, WeathersbyRSI15}. Femtosecond time-resolved photoelectron imaging has already been successfully applied to probe structural dynamics in molecules \cite{SuzukiARPC06, HolmegaardNature10, BollFarDiss14}, but its application to electron dynamics, which is much faster, is limited by temporal resolution. Recent progress in attosecond science makes it feasible to produce isolated attosecond pulses with energies reaching the XUV region \cite{ChiniNature14} that enable overcoming this temporal barrier \cite{VrakkingPCCP14}. Imaging molecular electron dynamics with high-energy photoelectrons generated by such XUV pulses would benefit not only from \r Angstrom spatial resolution, but also from a less complicated interpretation than that in the case of low-energy photoelectrons, where multiple scattering effects can be considerable \cite{SpringerBookAMO}.

The idea of using angle-resolved photoelectron distributions induced by single-photon ionization to image coherent electron dynamics has already been considered theoretically \cite{MignoletPRA12, KusJPCA13, PerveauxJPCA14, MignoletJPB14}. In these studies, the effect of the broad bandwidth of an ultrashort probe pulse has either been neglected \cite{MignoletPRA12, KusJPCA13, MignoletJPB14} or approximated \cite{PerveauxJPCA14}. However, we have earlier demonstrated that an accurate treatment within the quantum electrodynamics (QED) approach taking into account all transitions that can be induced by a broadband probe pulse is necessary for a correct description of ultrafast x-ray scattering from a nonstationary electronic system \cite{DixitPNAS12,DixitPRA14,PopovaGorelovaPRB15_1,PopovaGorelovaPRB15_2}. Therefore, since this should also be true for time-resolved photoelectron imaging, we apply the QED treatment to describe the interaction of a nonstationary electronic system with an ultrashort isolated probe pulse inducing single-photon ionization. Thereby, we not only take into account the effect of the broad bandwidth of the probe pulse, but we also accurately treat the evolution of the probed electronic system during the probe pulse.

The article is organized as follows. We provide a theoretical description of TRARPES in Sec.~\ref{SectionTDPP}, and illustrate in Sec.~\ref{Section_slow_indole} our results by calculating time- and angle-resolved photoelectron spectra for XUV-induced second ionization of a singly valence-ionized indole molecule (C$_8$H$_7$N) with coherent electron dynamics. In Sec.~\ref{Section_Fourier}, we show that the Fourier transform of the photoelectron angular distributions at a fixed photoelectron energy is given by a linear combination of autocorrelation functions of time-dependent Dyson orbitals. Finally, in Sec.~\ref{Section_fast_indole}, we study how electron dynamics during the action of the probe pulse affects time-resolved photoelectron spectra.

\section{Time-dependent photoelectron probability}
\label{SectionTDPP}

We consider an electronic system with Hamiltonian $\hat H_{\text{m}}$ with eigenstates $|\Phi^{N_{\el}}_I\ra$ and eigenenergies $E_{I}$, where a pump pulse has launched a coherent electronic wave packet at time $t_0$ by creating a coherent superposition of electronic eigenstates. Thus, the density operator of the electronic system starts evolving in time as
\begin{equation}
\hat\rho^m(t)=|\Psi(t)\ra\la\Psi(t)|,\label{den_op}
\end{equation}
where $\Psi(t)=\sum_IC_Ie^{-iE_I(t-t_0)}|\Phi^{N_{\el}}_I\ra$, $C_I$ are time-independent coefficients determined by the pump pulse and $N_{\el}$ is the number of bound electrons in the system after the pump pulse and before the probe pulse (atomic units are used throughout this paper). We investigate which information can be obtained from momentum distributions of photoelectrons emitted from this electronic system due to the interaction with an ultrashort photoionizing XUV probe pulse.

We describe the interaction between the electronic system and the photoionizing ultrashort probe pulse using a similar formalism to the one applied for the study of ultrafast x-ray scattering \cite{DixitPNAS12,DixitPRA14,PopovaGorelovaPRB15_1,PopovaGorelovaPRB15_2}. Since we assume that the probe pulse does not temporally overlap with the pump pulse, the interaction of the electronic system with the probe pulse can be considered independently from the pump pulse \cite{SantraPRA11}. The total Hamiltonian of the whole system, electronic system and light, is  \cite{Loudon}
\begin{equation}
\hat H = \hat H_{\text{m}}+\sum_{\mathbf{k},s}\omega_{\k,s}\hat a_{\k,s}^\dagger\hat a_{\k,s}+\hat H_{\text{int}},
\end{equation}
where $\hat a_{\k,s}^\dagger$ and $\hat a_{\k,s}$ are creation and annihilation operators of a photon in the $\k$, $s$ mode with energy $\omega_{\k}=|\k|c$,  $c$ is the speed of light. 
\begin{equation}
\hat H_{\text{int}}=
\frac1c\int d^3r\hat \psi^\dagger(\mathbf r)\lf(\hat{\mathbf A}(\mathbf r)\cdot\mathbf p\rt)\hat \psi(\mathbf r),\label{H_int}
\end{equation}
is the minimal coupling interaction Hamiltonian in the Coulomb gauge, where $\hat{\mathbf A}$ is the vector potential operator of the electromagnetic field, $\mathbf p$ is the canonical momentum of an electron, $\hat \psi^\dagger$ and $\hat \psi$ are electron creation and annihilation field operators. 

The probability of observing an emitted electron with momentum $\mathbf q$ within the density matrix formalism \cite{Mandel} is given by
\begin{equation}
P=\lim_{t_f\to+\infty}\Tr\lf [\hat O_{\mathbf q}\hat\rho_f(t_f)\rt]\label{Probab},
\end{equation}
where $\hat\rho_f(t_f)$ is the total density matrix of the electron system and the electromagnetic field at time $t_f$ after the action of the probe pulse.
The operator 
\begin{equation}
\hat O_{\mathbf q} =\sum_{\sigma}\hat c^\dagger_{\mathbf q,\sigma}\hat c_{\mathbf q,\sigma}\label{Oks}.
\end{equation}
describes the observation of a photoelectron with momentum $\mathbf q$ and spin $\sigma$, $\hat c_{\mathbf q,\sigma}$ ($\hat c_{\mathbf q,\sigma}^\dagger$) annihilates (creates) an electron with momentum $\mathbf q$ and spin $\sigma$. Evaluating $\hat\rho_f(t_f)$ within the first-order time-dependent perturbation theory using $H_{\text{int}}$ as the perturbation and applying the dipole approximation, we show in Appendix~\ref{AppendixTDPP} that the probability of observing an emitted electron due to the interaction with the probe pulse with duration $\tau_p$ and intensity $I_{\text{in}}(t) = I_0\,e^{-4\ln2[(t-t_p)/\tau_p]^2}$ is
%
%
\begin{align}
P =& \frac{\pi^2 \tau_p^2 I_0}{\ln2\omega_\i^2 c}\sum_{F,\sigma} \Bigg|\sum_{I}\nu_I\int d^3 r\phi_e^\dagger(\mathbf r, \mathbf q)\,(\boldsymbol\epsilon_\i\cdot\mathbf p)\label{Photoel_prob_general}\\
&\times\la\Phi_{F}^{N_{\el}-1}|\hat\psi(\mathbf r)|C_Ie^{-iE_I(t_p-t_0)}\Phi^{N_{\el}}_I\ra\Bigr|^2\nonumber,\\
\nu_I = &e^{-(\omega_\i-E_F^{N_{\el}-1} +E_{I}-\varepsilon_{e})^2\tau_p^2/(8\ln2)},\nonumber
%
\end{align}
where $|\Phi_F^{N_{\el}-1}\ra$ is a final state of the electronic system with $N_{\el}-1$ electrons, which by assumption does not interact with the emitted  photoelectron, $\phi_e$ is the photoelectron wave function, $\varepsilon_{e} = |\mathbf q|^2/2$ is the photoelectron energy and $\boldsymbol\epsilon_{\i}$ is the polarization vector of the probe pulse. 

Equation (\ref{Photoel_prob_general}) does not make use of the assumption that the probe-pulse duration is much shorter than the characteristic time scale of changes in the electron density, which we will refer to as the ``frozen-density approximation". As a consequence, the sum $\sum_IC_Ie^{-iE_I(t_p-t_0)}$ cannot be singled out due to the presence of the factor $\nu_I$ and the electronic wave packet state $\Psi(t_p)$ [cf.~Eq.~(\ref{den_op})] does not enter the expression for the photoelectron probability in this general case. Applying the frozen-density approximation allows substituting $E_I$ in the factors $\nu_I$ for the mean energy of the electronic wave packet $\la E \ra$. Then, the photoelectron probability is connected to $\Psi(t_p)$:
\begin{align}
P(\mathbf q,t_p) =&\frac{\pi^2 \tau_p^2 I_0}{\ln2\omega_\i^2 c} \sum_{F,\sigma}e^{-(\Omega_F -\varepsilon_{e})^2\tau_p^2/(4\ln2)}\label{Photoelprob-wp}\\
&\qquad\times\Bigl|\int d^3 r\phi_e^\dagger(\mathbf r, \mathbf q)\,(\boldsymbol\epsilon_\i\cdot\mathbf p)\phi_{F}^D(\mathbf r, t_p)\Bigr|^2,\nonumber
\end{align}
where
\begin{align}
\phi_{F}^D(\mathbf r, t_p) =& \la\Phi_{F}^{N_{\el}-1}|\hat\psi(\mathbf r)|\Psi(t_p)\ra\label{tdDyson_orbital}
\end{align}
is a time-dependent Dyson orbital connected to the wave packet state $\Psi(t_p)$ and, $\Omega_F = \omega_\i-E_F^{N_{\el}-1}+\la E \ra$. This expression is only valid as long as the maximal energy splitting among the eigenenergies of the electronic wave packet is negligible in comparison to the probe pulse bandwidth. As we will demonstrate in Sec.~\ref{Section_fast_indole}, the connection of the time-resolved photoelectron probability to the wave packet state $\Psi(t_p)$ breaks down otherwise. 

The expression for the time-resolved photoelectron probability in Eq.~(\ref{Photoelprob-wp}) is similar, but not equal to that derived in Ref.~\onlinecite{MignoletPRA12}. The difference is that we do not assume that ionization occurs only at the maximum of the probe pulse field. Indeed, we take into account the consequence of the broad probe-pulse bandwidth that gives rise to the exponential factor in Eq.~(\ref{Photoelprob-wp}), which is critical for a correct interpretation of time-resolved photoelectron spectra, as will be shown in the next section.

According to Eq.~(\ref{Photoelprob-wp}), time-dependent photoelectron spectra at each emission angle consist of a series of photoelectron peaks centered at energies $\Omega_F$ corresponding to a transition to a final state $F$. The strength of the peaks depends on the emission angle and on the probe-pulse arrival time; however, the strength of some peaks is time-independent, as will be shown below. The centers of the peaks, $\Omega_F$, are angle- and time-independent. This conclusion contradicts the intuitive picture that, since, in a stationary measurement, electron binding energies depend on the local chemical environment, photoelectron peaks would shift following the time-dependent electron density in a time-resolved measurement.

Let us consider the $F$-th term in the sum over final states in Eq.~(\ref{Photoelprob-wp}) in more detail. It is proportional to $\sum_I|C_I|^2|\widetilde\phi_{FI}^D(\mathbf q)|^2+\sum_{I\neq K}C_IC_K^*e^{-i(E_I-E_K)(t_p-t_0)} \widetilde\phi_{FI}^D(\mathbf q) \widetilde\phi_{FK}^{D*}(\mathbf q)$, where $\widetilde\phi_{FI(FK)}^D(\mathbf q)=\int d^3r\phi_e^\dagger(\boldsymbol\epsilon_\i\cdot\mathbf p)\la\Phi_F^{N_{\el}-1}|\hat\psi| \Phi_{I(K)}^{N_{\el}}\ra$. The first sum over $I$ provides a time-independent and the second sum over $I\neq K$ provides a time-dependent contribution to the time-resolved photoelectron probability. The $I,K$-th term in the second sum is nonzero only if both functions $\la\Phi_F^{N_{\el}-1}|\hat\psi| \Phi_{I}^{N_{\el}}\ra$ and $\la\Phi_F^{N_{\el}-1}|\hat\psi| \Phi_{K}^{N_{\el}}\ra$ are nonzero, {\it i.e.}, a transition to a final state $\Phi_F^{N_{\el}-1}$ by the emission of a photoelectron is allowed for both eigenstates $\Phi_I$ and $\Phi_K$. This means that if the transition to a final state $\Phi_F^{N_{\el}-1}$ is possible only from a single eigenstate involved in the electronic wave packet, the $F$-th term in Eq.~(\ref{Photoelprob-wp}) is nonzero, but time-independent. The strength of the peaks corresponding to such transitions is also nonzero and does not depend on the probe-pulse arrival time.


We apply the plane-wave approximation to the photoelectron wave function 
\begin{align}
|\phi_e\ra=|q_e\ra = \frac{1}{\sqrt{(2\pi)^3}}e^{i\mathbf q\cdot \mathbf r}\chi_e(\sigma),
\end{align}
where $\chi_e(\sigma)$ is the photoelectron spin state, for the further analysis of the time-dependent photoelectron probability. An advantage of this approximation is that it considerably simplifies the interpretation of the angle-resolved photoelectron spectra, since the integral in Eq.~(\ref{Photoelprob-wp}) reduces to
\begin{align}
&\int d^3 r q_e^\dagger\,(\boldsymbol\epsilon_\i\cdot\mathbf p)\phi_{F}^D  \\
&\qquad=  \chi_e^\dagger(\sigma)\frac{(\boldsymbol\epsilon_\i\cdot\mathbf q)}{\sqrt{(2\pi)^3}}\int d^3r e^{-i\mathbf q\cdot \mathbf r}\phi_{F}^D(\mathbf r,t_p)\nonumber.
\end{align}
Although the plane-wave approximation can yield incorrect results in the case of some molecules or experimental conditions \cite{BurkeJChPh76}, it has been shown that this approximation is appropriate for the calculation of angle-resolved photoemission spectra from $\pi$-orbitals of planar molecules consisting of light atoms at small angles between $\boldsymbol\epsilon_\i$ and $\mathbf q$ \cite{PuschnigScience09,PuschnigChapter13,PuschnigJESRP15}. Still, our main motivation to apply the plane-wave approximation is to provide an example of an analysis of time- and angle-resolved photoelectron spectra within the most straightforward approximation to the photoelectron wave function that leads to the following result:
\begin{align}
P(\mathbf q,t_p) =&\frac{ \tau_p^2 I_0|\boldsymbol\epsilon_\i\cdot\mathbf q|^2}{8\pi\ln2\omega_\i^2 c} \sum_{F,\sigma}e^{-(\Omega_F -\varepsilon_{e})^2\tau_p^2/(4\ln2)}\label{Probab_wp}\\
&\quad\times\Bigl|\chi_e^\dagger(\sigma)\int d^3r e^{-i\mathbf q\cdot \mathbf r}\phi_{F}^D(\mathbf r,t_p)]\Bigr|^2.\nonumber
\end{align}

\section{Time- and angle-resolved photoelectron spectra of indole}
\label{Section_slow_indole}

\begin{figure}[t]
\includegraphics[width=0.5\textwidth]{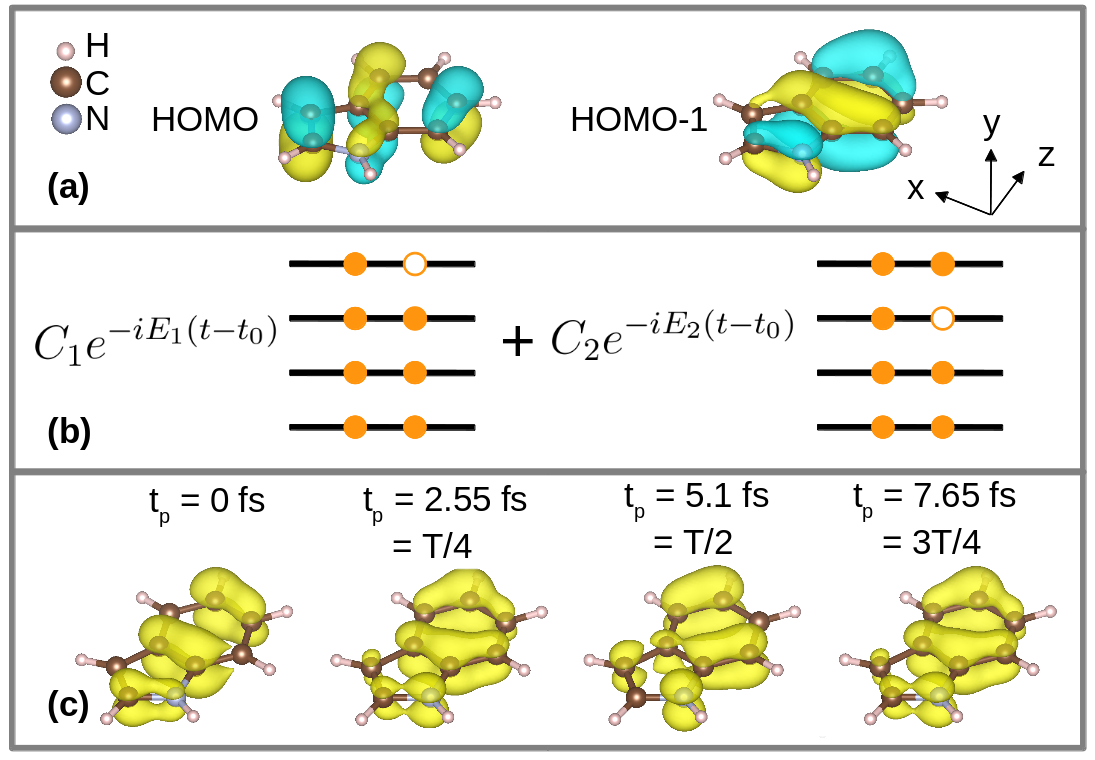}
  \caption{(a) Isosurfaces of the amplitudes of the HOMO and HOMO-1 orbitals of indole calculated with the software package MOLCAS \cite{KarlstromPSSD03}. (b) Schematic representation of the electronic state of indole after the pump pulse arriving at time $t_0$, which is the superposition of two states with an electron hole in HOMO and HOMO-1 orbitals. (c) Electron hole density at time $0$, $T/4$, $T/2$ and $3 T/4$, where $T=2\pi/(E_2-E_1)$. Electron hole densities coincide at $T/4$ and $3 T/4$. The orbitals and electron hole densities are vizualized using the VESTA software \cite{MommaJAC11}.}
\label{Fig_AfterPump}
\end{figure}

We apply our formalism to the calculation of angle- and time-resolved photoelectron spectra of an indole molecule with coherent electron dynamics in valence orbitals. A common experimental technique to trigger coherent electron dynamics in a molecule in a controlled way is to create an electron hole in a superposition of valence orbitals by a broadband photoionizing pump pulse, as has been demonstrated in Refs.~\onlinecite{SmirnovaNature09, GoulielmakisNature10, SansoneNature10, HaesslerNature10, TzallasNature11, CalegariScience14}. Thus, we assume that a broadband photoionizing pump pulse launched an electronic wave packet in indole by creating an electron hole in a superposition of the HOMO (highest occupied molecular orbital) and HOMO-1 orbitals at time $t_0$ [see Fig.~\ref{Fig_AfterPump}(a) and (b)]. Then, the electronic state of the indole molecular cation after the interaction with the pump pulse evolves in time: 
\begin{align}
|\Psi(t)\ra= C_1e^{-iE_1(t-t_0)}|\Phi_{H}^{\text{ion}+}\ra+C_2e^{-iE_2(t-t_0)}|\Phi_{H-1}^{\text{ion}+}\ra,\label{wp-indole}
\end{align}
where $|\Phi_{H}^{\text{ion}+}\ra$ is an electronic state with an electron hole in the HOMO, and $|\Phi_{H-1}^{\text{ion}+}\ra$ is an electronic state with an electron hole in the HOMO-1. We obtain the energies $E_1$ and $E_2$ of $|\Phi_{H}^{\text{ion}+}\ra$ and $|\Phi_{H-1}^{\text{ion}+}\ra$ within Koopmans' theorem \cite{KoopmansPhysica34}. $C_1$ and $C_2$ are time-independent complex coefficients such that $|C_1|^2+|C_2|^2=1$, which are determined by the pump process.  In our study, we concentrate on the information that the TRARPES probe technique can provide about coherent electronic dynamics independently from how it has been excited. Therefore, we do not describe the pump process and just randomly choose $C_1$ and $C_2$ coefficients. Nevertheless, a few general remarks must be made.

In order to trigger the coherent electronic wave packet that we consider, it is necessary to choose the bandwidth and photon energy of the pump pulse such that the bandwidth is larger than the energy splitting between the HOMO and HOMO-1 orbitals, but the photon energy is not sufficiently large to ionize the HOMO-2 orbital. From the calculation of the electronic structure of neutral indole within the Hartree-Fock approach using the ab initio quantum chemistry software package MOLCAS \cite{KarlstromPSSD03} with the cc-pVDZ basis set \cite{DunningJChPh89, EMSLDataBase}, we obtained indole HOMO, HOMO-1 and HOMO-2 binding energies of 7.8 eV, 8.2 eV and 10.5 eV, respectively. Thus, the energy splitting between the HOMO and HOMO-1 is small enough and the splitting between the HOMO-1 and HOMO-2 is large enough to identify suitable pump pulse parameters.

After the pump process, the electron density starts oscillating with the period $T=2\pi/(E_2-E_1)\approx$ 10.2 fs and with the time-dependent part being proportional to $\cos[(E_1-E_2)(t_p-t_0) - \varphi_{12}]$, where $\varphi_{12}=\operatorname{arg}(C_1C_2^*)$. Thus, there are two time points during the density oscillation period, at which the time-dependent electron densities are identical (this is always true when only two electronic states are involved in electron dynamics). We define the zero point on the time scale at a time equal to $t_0 + \phi_{12}/(E_1-E_2)$. With this definition, the time-dependent part of the electron density is proportional to $\cos[(E_1-E_2)t_p]$ and the times when densities coincide are $t_p=T/4$ and $t_p=3T/4$ [see Fig.~\ref{Fig_AfterPump}(c)]. We assume, for simplicity, that nuclear motion is negligible on the 10 fs time scale and that the atoms have not had time to move since the interaction of the molecule with the pump pulse.

\begin{figure}[t]
\centering
\subfloat[]{\includegraphics{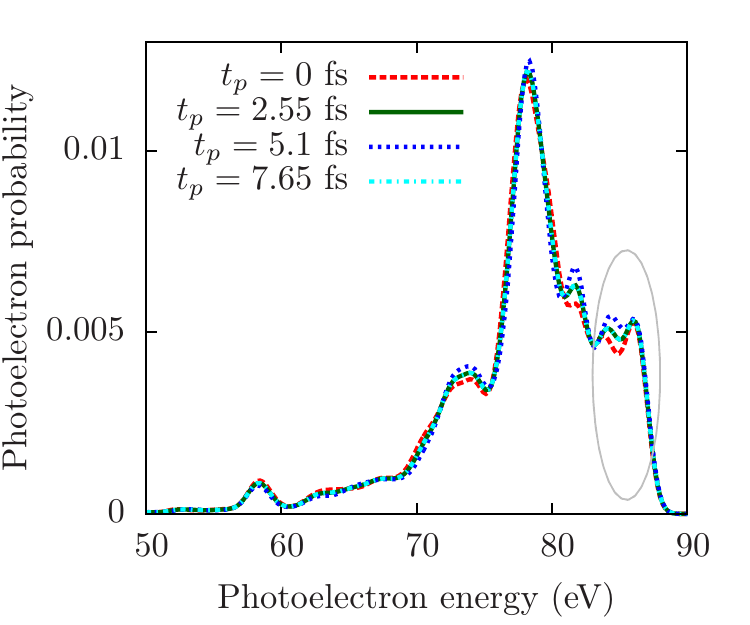}}\\
\subfloat[]{\includegraphics{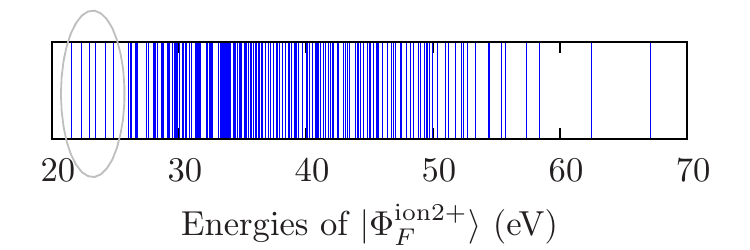}}
\caption{(a) Angle-averaged photoelectron spectra at different probe-pulse arrival times. (b) Energies of possible final dicationic states of indole.}
\label{Fig_spectra}
\end{figure}

We assume that a $y$-polarized probe pulse of 1 fs duration, 100 eV photon energy and $10^{12}$ W/cm$^2$ intensity photoionizes the indole molecular cation at time $t_p$, and an emitted electron of momentum $\mathbf q$ is detected. Thereby, the probe pulse creates a second electron hole in indole. We have verified that the 1 fs probe-pulse duration is short enough to satisfy the frozen-density approximation. The high photon energy of the probe pulse is chosen to gain a high spatial resolution (photoelectron de Broglie wavelength of the order of 1 \r A) and minimize the orthogonalization correction to the plane-wave approximation, as will be shown below. The polarization vector of the pulse is perpendicular to the nodal plane of indole (cf.~Fig.~\ref{Fig_AfterPump}).


In order to calculate the photoelectron probability with Eq.~(\ref{Probab_wp}), one has to determine the states $\Phi_F^{N_{\el}-1}$ [see Eq.~(\ref{tdDyson_orbital})], which are dicationic states of indole, $\Phi_F^{\text{ion}2+}$ ($N_{\el}$ is by definition the number of electrons after the interaction with the pump pulse). We perform this calculation with the software package MOLCAS \cite{KarlstromPSSD03} applying the configuration interaction method within the two-hole configuration space \cite{CImethod}. Within this method, a doubly-ionized electronic state with total spin $S$ and spin projection $M_S$ is represented as $\Phi_F^{\text{ion}2+} = \sum_{i\ge 0,j\ge i}\alpha_{ij}^{F,S}\Phi_{H-i,H-j}^{\text{ion}2+,S(M_S)}$, where 
\begin{align}
&\Phi_{H-i,H-j}^{\text{ion}2+,0(0)} = \Big[\Phi_{H-i\uparrow,H-j\downarrow}^{\text{ion}2+}-\Phi_{H-i\downarrow,H-j\uparrow}^{\text{ion}2+}\Big]/\sqrt{2}\\
&\Phi_{H-i,H-j}^{\text{ion}2+,1(+1)} = \Phi_{H-i\uparrow,H-j\uparrow}^{\text{ion}2+}\\
&\Phi_{H-i,H-j}^{\text{ion}2+,1(0)} = \Big[\Phi_{H-i\uparrow,H-j\downarrow}^{\text{ion}2+}+\Phi_{H-i\downarrow,H-j\uparrow}^{\text{ion}2+}\Big]/\sqrt{2}\\
&\Phi_{H-i,H-j}^{\text{ion}2+,1(-1)} = \Phi_{H-i\downarrow,H-j\downarrow}^{\text{ion}2+}
\end{align}
for $j\neq i$. $\Phi_{H-i\uparrow(\downarrow),H-j\uparrow(\downarrow)}^{\text{ion}2+}$ is a configuration state function with two electron holes in orbitals HOMO-$i$ and HOMO-$j$ of neutral indole, the arrows denoting the spin of the remaining electrons in these orbitals. $\Phi_{H-i,H-i}^{\text{ion}2+,0(0)}$ is a configuration state function with two electron holes in the HOMO-$i$ orbital. It follows from the Pauli principle that $\alpha_{ii}^{F}=0$ for $S=1$. Triplet states that differ only by spin projections are different final states, but with equal energy and the same corresponding expansion coefficients $\alpha_{ij}^{F,S}$. Although spin-unresolved results do not depend on the spin of the indole molecular cation, we assume, for convenience, that its projection $M_S$ was equal to +1/2 before the ionization by the probe pulse. Then, Dyson orbitals corresponding to final states with $S=0$, $M_S=0$ and $S=1$, $M_S=0$ have a spin equal to $-1/2$ and Dyson orbitals corresponding to final states with $S=1$, $M_S=1$ have a spin equal to $+1/2$. The contribution from Dyson orbitals corresponding to final triplet states $F_1$ and $F_2$ that differ only by spin projection are proportional to each other: $\chi_{+1/2}^\dagger\phi^D_{F_1}(\mathbf r,t_p) = \sqrt{2}\chi_{-1/2}^\dagger\phi^D_{F_2}(\mathbf r,t_p)$, where the spin projection of $F_1$ is +1 and of $F_2$ is 0. A transition to a final triplet state with $M_S=-1$ is not possible.



\begin{table}
\centering
\begin{tabular}{| c | c | c | c |}
\hline
$F$ &$E_F^{\text{ion}2+}$ (eV)& $\Omega_F$ (eV)& $\Phi_F^{\text{ion}2+}$ \\
\hline & & & \\[-1.5ex]
1 & 21.6 & 86.4 & $-0.9\Phi_{H,H}^{\text{ion}2+,0(0)}$ \\ [0.5ex]
\hline & & & \\[-1.5ex]
2 & 21.6 & 86.4 & $-\Phi_{H,H-1}^{\text{ion}2+,1(+1)}$  \\[0.5ex]
\hline & & & \\[-1.5ex]
3 & 21.6 & 86.4 & $-\Phi_{H,H-1}^{\text{ion}2+,1(0)}$  \\[0.5ex]
\hline & & & \\[-1.5ex]
4 & 22.3 & 85.7 & $-0.9\Phi_{H,H-1}^{\text{ion}2+,0(0)}$ \\[0.5ex]
\hline & & & \\[-1.5ex]
5 & 22.9 & 85.1 & $0.9\Phi_{H,H-2}^{\text{ion}2+,1(+1)}$ \\[0.5ex]
\hline & & & \\[-1.5ex]
6 & 22.9 & 85.1 & $0.9\Phi_{H,H-2}^{\text{ion}2+,1(0)}$ \\[0.5ex]
\hline & & & \\[-1.5ex]
7 & 23.4 & 84.6 & $0.9\Phi_{H-1,H-1}^{\text{ion}2+,0(0)}$ \\[0.5ex]
\hline & & & \\[-1.5ex]
8 & 24.2 & 83.8 & $0.8\Phi_{H-1,H-2}^{\text{ion}2+,1(+1)}-0.5\Phi_{H,H-3}^{\text{ion}2+,1(+1)}$ \\[0.5ex]
\hline & & & \\[-1.5ex]
9 & 24.2 & 83.8 & $0.8\Phi_{H-1,H-2}^{\text{ion}2+,1(0)}-0.5\Phi_{H,H-3}^{\text{ion}2+,1(0)}$ \\[0.5ex]
\hline & & & \\[-1.5ex]
10 & 24.8 & 83.2 & $-0.8\Phi_{H-1,H-2}^{\text{ion}2+,0(0)}-0.5\Phi_{H,H-3}^{\text{ion}2+,0(0)}$ \\[0.5ex]
\hline
\end{tabular}
\caption{Expansion of possible final states $\Phi_F^{\text{ion}2+}$ with energies $E_F^{\text{ion}2+}$ below 26 eV in terms of $\Phi_{H-i,H-j}^{\text{ion}2+,S(M_S)}$ (only terms with coefficients $|\alpha_{ij}^{F,S}|\ge0.4$ are shown). $\Omega_F$ is the center of the Gaussian peak contributed by the corresponding final state to the photoelectron spectrum.}
\label{table_FinalStates}
\end{table}

\begin{figure*}
\centering
\includegraphics[width = \textwidth]{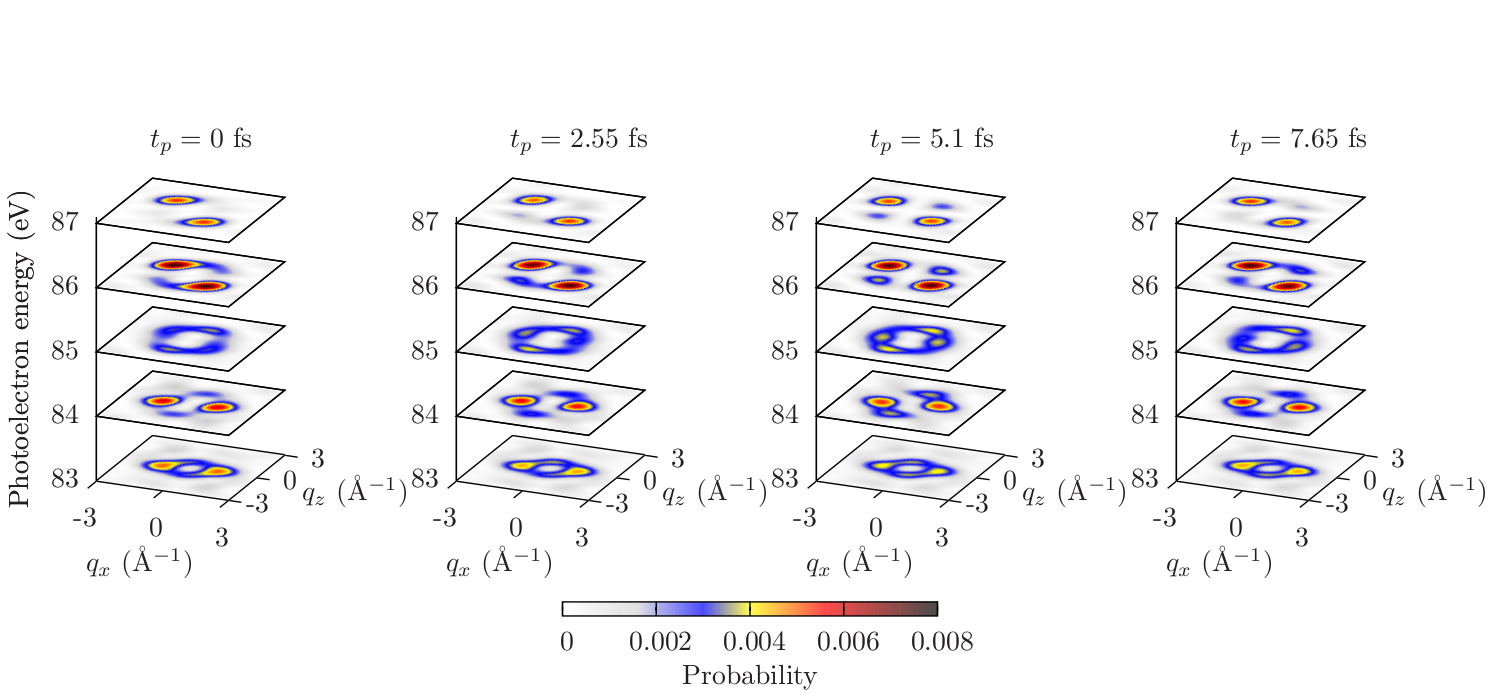}
\caption{Angle-resolved photoelectron spectra generated by the probe pulse arriving at time $t_p$. Each $q_xq_z$ plane at the corresponding photoelectron energy $\varepsilon_e$ is a projection of the semisphere with radius $|\mathbf q| = \sqrt{2\varepsilon_e}$ at $q_y>0$, where the color of the $\mathbf q$ point on the semisphere corresponds to the probability of detecting an electron with momentum $\mathbf q$.}
\label{Fig_arspectra}
\end{figure*}

\subsection{Angle-averaged photoelectron spectra}

First, we study angle-averaged photoelectron spectra of indole shown in Fig.~\ref{Fig_spectra}(a) for different probe-pulse arrival times. Although the changes in the electron hole density are considerable [see Fig.~\ref{Fig_AfterPump}(c)], the changes in the angle-averaged spectra with time are quite small. The spectra at $t_p=2.55$ fs and $t_p=7.65$ fs, when the electron densities coincide, are identical.

Figure \ref{Fig_spectra}(b) shows energies $E_F^{\text{ion}2+}$ of possible final dicationic states of indole, {\it i.e.}, final states for which either $\la\Phi_F^{\text{ion}2+}|\hat\psi(\mathbf r)|\Phi_{H}^{\text{ion}+}\ra\neq0$ or $\la\Phi_F^{\text{ion}2+}|\hat\psi(\mathbf r)|\Phi_{H-1}^{\text{ion}+}\ra\neq0$. A contribution to the photoelectron spectra corresponding to each of these states is a Gaussian peak centered at a photoelectron energy of $\Omega_F=108\text{ eV}-E_F^{\text{ion}2+}$. The peaks are approximately $ 2$ eV broad for $\tau_p=1$ fs. Therefore, the spectral contributions of transitions to final states with energy-differences smaller than 2 eV are indistinguishable, and the spectrum is formed by several possible final states at almost every photoelectron energy, since the dicationic states are energetically close to each other [see Fig.~\ref{Fig_spectra}(b)].



The effect of transitions to various final states on the photoelectron spectra may be understood best by considering the spectral region outlined with the gray oval in Fig.~\ref{Fig_spectra}(a). It is formed due to transitions to the final states highlighted with the gray oval in Fig.~\ref{Fig_spectra}(b). The energies of these final states and the corresponding expansion in terms of the configuration state functions are shown in Table \ref{table_FinalStates}. As discussed in the previous section, only those transitions provide a time-dependent contribution to the photoelectron probability, for which both functions $\la\Phi_F^{\text{ion}2+}|\hat\psi(\mathbf r)|\Phi_{H}^{\text{ion}+}\ra$ and $\la\Phi_F^{\text{ion}2+}|\hat\psi(\mathbf r)|\Phi_{H-1}^{\text{ion}+}\ra$ are nonzero. Since the dicationic states 2, 3, 4, 8, 9 and 10 involve configuration state functions with electron holes in both HOMO and HOMO-1 orbitals, the transition to these states is possible from both eigenstates of the electronic wave packet. A transition to the dicationic states 1, 5 and 6 is possible only from the eigenstate $|\Phi_{H}^{\text{ion}+}\ra$, and a transition to the state 7 only from the eigenstate $|\Phi_{H-1}^{\text{ion}+}\ra$. Thus, the transitions to the states 2, 3, 4, 8, 9 and 10 provide both time-dependent and time-independent contributions; and the transitions to the states states 1, 5, 6 and 7 provide only time-independent contributions. 


Let us compare the configuration interaction method to a simplified picture where the two electron holes just occupy neutral orbitals of indole, {\it i.e.}, each dicationic state involves just a single configuration state function. In this case, only transitions to the two states with the configurations $\Phi_{H,H-1}^{\text{ion2+},0}$ and $\Phi_{H,H-1}^{\text{ion2+},1}$ would provide a time-dependent contribution to the spectrum. Transitions to other dicationic states would be possible only from one of the eigenstates involved in the dynamics. Thus, the photoelectron spectra would vary with respect to the probe-pulse arrival time just around the binding energies of $\Phi_{H,H-1}^{\text{ion2+},S}$, and all other spectral regions would be time-independent. This demonstrates that electron correlations in the remaining ion are crucial for understanding TRARPES spectra.

Since the angle-averaged photoelectron spectra do not change noticeably with time, they are not a good observable for electron dynamics. Therefore, let us consider the time evolution of angle-resolved photoelectron spectra.

\subsection{Angle-resolved photoelectron spectra}

\begin{figure}
\centering
\includegraphics{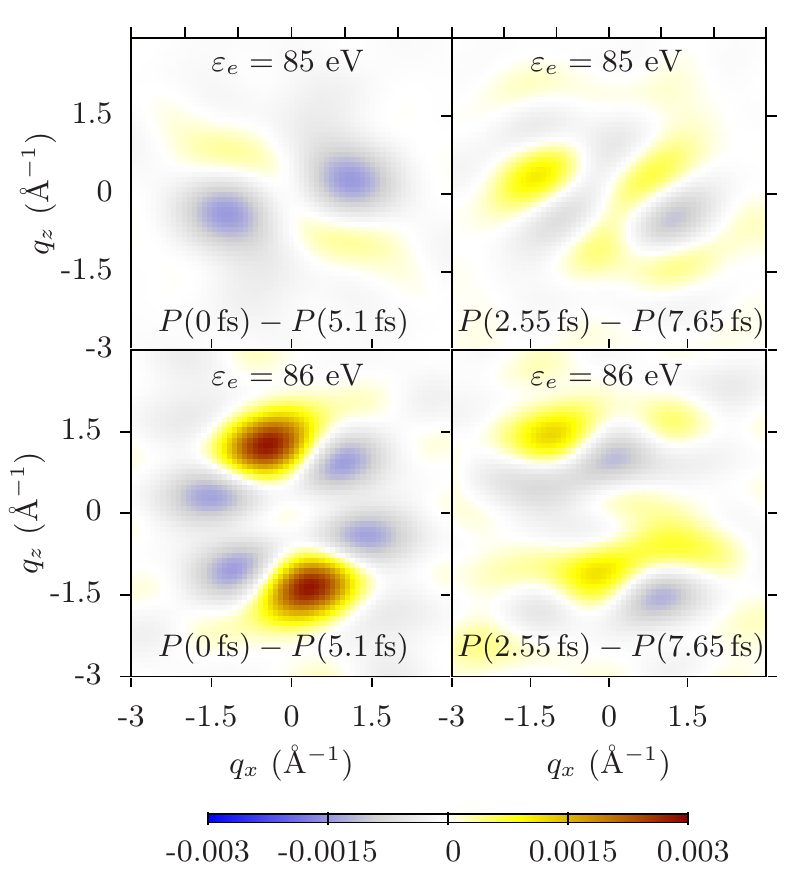}
\caption{The differences between the angle-resolved probability distributions at times $t_p=0$ fs and $t_p=5.1$ fs, $P(0\text{ fs})-P(5.1 \text{ fs})$, and at times $t_p=2.55$ fs and $t_p=7.65$ fs, $P(2.55\text{ fs})-P(7.65\text{ fs})$. The differences are shown for energies $\varepsilon = 85$ eV and 86 eV.}
\label{Fig_arspectra_t2-t4}
\end{figure}

Figure \ref{Fig_arspectra} shows angle-resolved photoelectron distributions at the photoelectron energies outlined with the gray oval in Fig.~\ref{Fig_spectra}(a), at different probe-pulse arrival times $t_p$. The $q_xq_z$ plane at a photoelectron energy $\varepsilon_e$ in Fig.~\ref{Fig_arspectra} is a projection of a semisphere with a radius $|\mathbf q|=\sqrt{2\varepsilon_e}$ constrained by $q_y>0$. The color of a $\mathbf q$ point on this semisphere corresponds to the probability to detect an electron with momentum $\mathbf q$. The angle-resolved distributions are symmetric with respect to the nodal plane of the planar molecule, {\it i.e.}, the semispheres are identical at opposite $q_y$. The angle between the probe-pulse polarization vector along the $y$ direction and the most probable directions in which electrons at these energies are emitted, is rather small. The seven Dyson orbitals that according to Table \ref{table_FinalStates} determine the spectra at 83-86 eV are $\pi$ orbitals. Thus, the conditions of the applicability of the plane-wave approximation mentioned in the previous section are satisfied. The angle-resolved photoelectron spectra calculated with the photoelectron wave function $|\phi_e\ra= |q_e\ra - \sum_\alpha|\phi_\alpha\ra\la\phi_\alpha|q_e\ra$ \cite{RabalaisJChPh74}, which is a plane  wave orthogonalized to the occupied molecular orbitals of the cation, $\phi_\alpha (\mathbf r)$, are shown in Fig.~\ref{Fig_arpes_ortho} in Appendix \ref{Appendix_Ortho} and are almost identical to those shown in Fig.~\ref{Fig_arspectra}.

The changes in the angular distributions of the photoelectron spectra in time (Fig.~\ref{Fig_arspectra}) are much larger in comparison to the angle-averaged spectra (Fig.~\ref{Fig_spectra}(a)). Thus, for detecting electron dynamics in a molecule, angle-resolved photoelectron spectra are much more powerful than angle-averaged ones. The difference between the photoelectron spectra are maximal at times $t_p=0$ fs and $t_p=5.1$ fs (0 and $T/2$), approaching 40 \% at $\varepsilon_e=85$ eV and 86 eV (cf.~Fig.~\ref{Fig_arspectra_t2-t4}). The angle-resolved photoelectron spectra also do not coincide at times $t_p=2.55$ fs and $t_p=7.65$ fs ($T/4$ and $3T/4$), although the electron densities (see Fig.~\ref{Fig_AfterPump}) and the angle-averaged spectra [Fig.~\ref{Fig_spectra}(a)] are identical at these times. For example, the difference between the angle-resolved photoelectron distributions at $t_p=2.55$ fs and $t_p=7.65$ fs is about 25 \% at $\varepsilon_e=85$ eV and 86 eV, as shown in Fig.~\ref{Fig_arspectra_t2-t4}. 


In order to understand why the angle-resolved photoelectron spectra do not follow the electron density, but the angle-averaged spectra do, we rewrite Eq.~(\ref{Probab_wp}) in the following form
\begin{align}
& P(\mathbf q,t_p)\propto |\boldsymbol\epsilon_\i\cdot\mathbf q|^2\sum_{F,\sigma}e^{-\frac{(\Omega_F -\varepsilon_{e})^2\tau_p^2}{4\ln2}}\int d^3r_1\int d^3r_2\nonumber\\
&\times\Big\{\cos(\mathbf q\cdot[\mathbf r_1-\mathbf r_2])\Re\la\Psi(t_p)|\hat G_F(\mathbf r_1,\mathbf r_2) |\Psi(t_p)\ra \label{Probab_wp_Re_Im}\\
&\quad+ \sin(\mathbf q\cdot[\mathbf r_1-\mathbf r_2])\Im \la\Psi(t_p)|\hat G_F(\mathbf r_1,\mathbf r_2) |\Psi(t_p)\ra \Big\}\nonumber\\
&\hat G_F(\mathbf r_1,\mathbf r_2)=\hat\psi^\dagger(\mathbf r_2)|\Phi_F^{N_{\el}-1}\ra\la\Phi_F^{N_{\el}-1}|\hat\psi(\mathbf r_1)\nonumber
\end{align}
The first term in the curly braces has a similar time dependency as the electron density does
\begin{align}
\rho(\mathbf r,t_p)&=\la\Psi(t_p)|\hat\psi^\dagger(\mathbf r)\hat\psi(\mathbf r)|\Psi(t_p)\ra\\
&=\Re\la\Psi(t_p)|\hat\psi^\dagger(\mathbf r)\hat\psi(\mathbf r)|\Psi(t_p)\ra\nonumber.
\end{align}
In the scenario considered, the time-dependent part of the electron density and of the first term in the curly braces are each proportional to $\cos([E_1-E_2]t_p)$ [see the discussion of Fig.~\ref{Fig_AfterPump}(c)] and, thus, their time evolutions coincide. However, their time evolutions may not coincide in a general case, when more than two electronic states are involved in the electron dynamics. 

Since $\hat G_F(\mathbf r_1,\mathbf r_2)\neq \hat G_F^\dagger(\mathbf r_1,\mathbf r_2)$ for $\mathbf r_1\neq\mathbf r_2$, the second term in the curly braces in Eq.~(\ref{Probab_wp_Re_Im}) is nonzero and provides an additional time-dependent contribution, which is different from that of the electron density. In the case of the scenario considered, this term is proportional to $\sin([E_1-E_2]t_p)$. Therefore, angle-resolved photoelectron spectra do not follow the time evolution of the electron density. A similar effect that a measured signal from an electronic wave packet does not follow the time-dependent electron density has been shown for time-resolved resonant \cite{PopovaGorelovaPRB15_1,PopovaGorelovaPRB15_2} and nonresonant \cite{DixitPNAS12} x-ray scattering patterns as well as time-resolved electron diffraction patterns \cite{ShaoPRA13}. 

The angular average of the second term in the curly braces in Eq.~(\ref{Probab_wp_Re_Im}) is zero, since $\int_{0}^{2\pi}d\alpha\int_{-1}^1d\cos\beta|\boldsymbol\epsilon_\i\cdot\mathbf q|^2\sin(\mathbf q\cdot[\mathbf r_1-\mathbf r_2])=0$, where $\alpha$ and $\beta$ are polar angles in $\mathbf q$ space. Therefore, the angle-averaged spectra are determined only by the first term in the curly braces in Eq.~(\ref{Probab_wp_Re_Im}) and their time evolution follows the electron density. The contributions of the two terms in Eq.~(\ref{Probab_wp_Re_Im}) can be distinguished by performing a Fourier analysis, as will be shown in the next section.

\section{Fourier analysis of TRARPES}
\label{Section_Fourier}

In this Section, we perform a Fourier transform of the photoelectron angular distributions at a fixed photoelectron energy $\varepsilon_e=q_0^2/2$, \textit{i.e.,} of the spherical surfaces in $\mathbf q$ space of a fixed radius $q_0$, the projections of which are shown in Fig.~\ref{Fig_arspectra}. Performing a Fourier transform this way has several advantages. As we will show in this Section, first, one can concentrate on photoelectron angular distributions only at high photoelectron energies and, thereby, gain a high spatial resolution. Second, only few Dyson orbitals have to be considered for the interpretation of a Fourier transform at a fixed photoelectron energy, whereas a much larger number of Dyson orbitals contribute in the case of a Fourier transform involving an integration over $q_0$ of a whole angle-resolved photoelectron spectrum. Third, one can choose a photoelectron energy at which the changes of the angular distributions in time are much larger than at other energies, and gain a much better contrast than in the case of energy-unresolved photoelectron distributions.

The Fourier transform of the angular distributions of the photoelectron probability at a fixed photoelectron energy $\varepsilon_e=q_0^2/2$ is performed according to the following equation
\begin{align}
&\mathcal F(\mathbf r,\varepsilon_e,t_p) =\int d^3 q \frac{P(\mathbf q,t_p)}{|\boldsymbol\epsilon_\i\cdot\mathbf q|^2}e^{i \mathbf q \cdot\mathbf r}\delta(|\mathbf q|-q_0)\label{F_2D}\\
 &=\frac{ \tau_p^2 I_0 \varepsilon_e}{\ln2\omega_\i^2 c} \sum_{F,\sigma} e^{-(\Omega_F -\varepsilon_{e})^2\tau_p^2/(4\ln2)}(A[\phi_F^D(t_p)]*s)(\mathbf r),\nonumber
\end{align}
where $\delta(|\mathbf q|-q_0)$ is the Dirac Delta function. $A[\phi_{F}^D(t_p)] = \int d^3 r'{\phi_{F}^D}^\dagger(\mathbf r'-\mathbf r,t_p)\phi_{F}^D(\mathbf r',t_p) $ is the autocorrelation function of $\phi_{F}^D(\mathbf r,t_p)$, $(A[\phi_F^D(t_p)]*s)(\mathbf r)=\int d^3r'A(\mathbf r')s(\mathbf r-\mathbf r')$ denotes the convolution of the autocorrelation function $A[\phi_F^D(t_p)]$ with the function $s(\mathbf r)=\operatorname{sinc}(q_0r)$, where $r = |\mathbf r|$. The function $\operatorname{sinc}(q_0r)$ has a broad peak centered at $r=0$ and much weaker side peaks. The width of the central peak, which is proportional to $1/q_0$, determines the spatial resolution of the measurement, which reflects the fact that the spatial resolution is limited by the photoelectron de Broglie wavelength. At a probe-pulse photon energy of 100 eV, the resolution of the measurement is about 1 \r A. In comparison, 1 \r A spatial resolution using x-ray scattering can only be achieved using a hard x-ray pulse with a photon energy higher than 10 keV. 

\begin{figure}
\centering
\subfloat[]{\includegraphics[width = 0.45\textwidth]{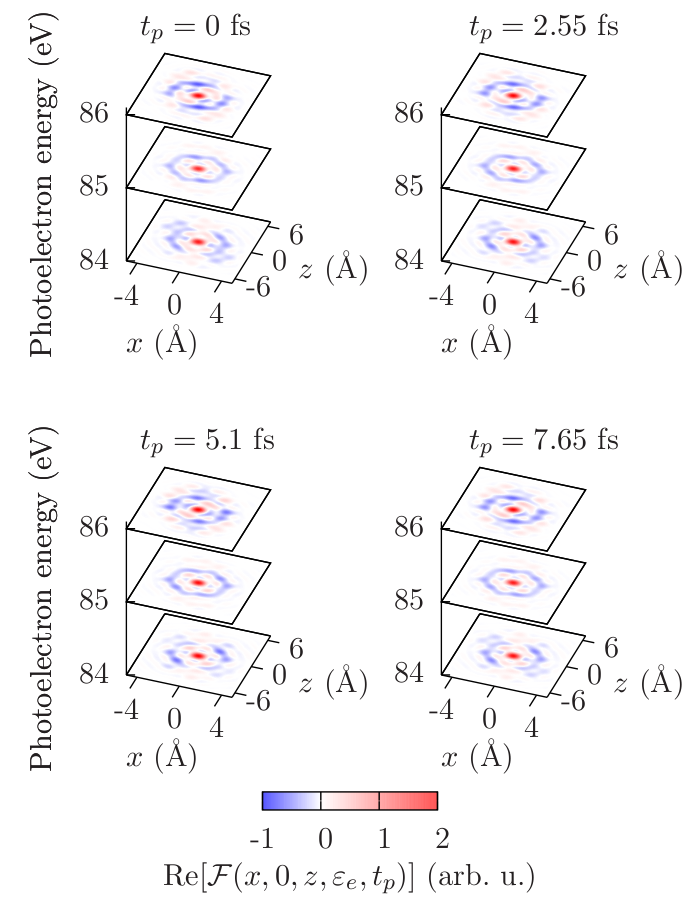}}\\
\subfloat[]{\includegraphics[width = 0.45\textwidth]{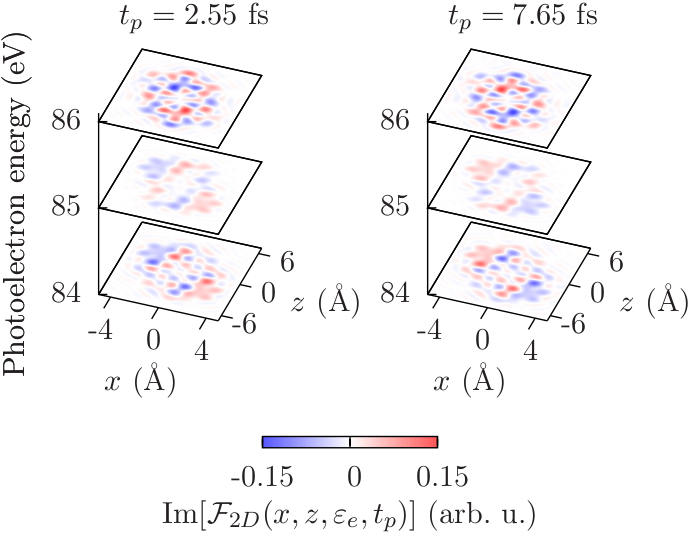}}
\caption{The real (a) and imaginary (b) parts of the Fourier transforms of the angle-resolved spectra in Fig.~\ref{Fig_arspectra}(a) according to Eq.~(\ref{F_2D}) at different times $t_p$.}
\label{Fig_F2D}
\end{figure}

\begin{figure}
\centering
\subfloat[]{\includegraphics[width = 0.45\textwidth]{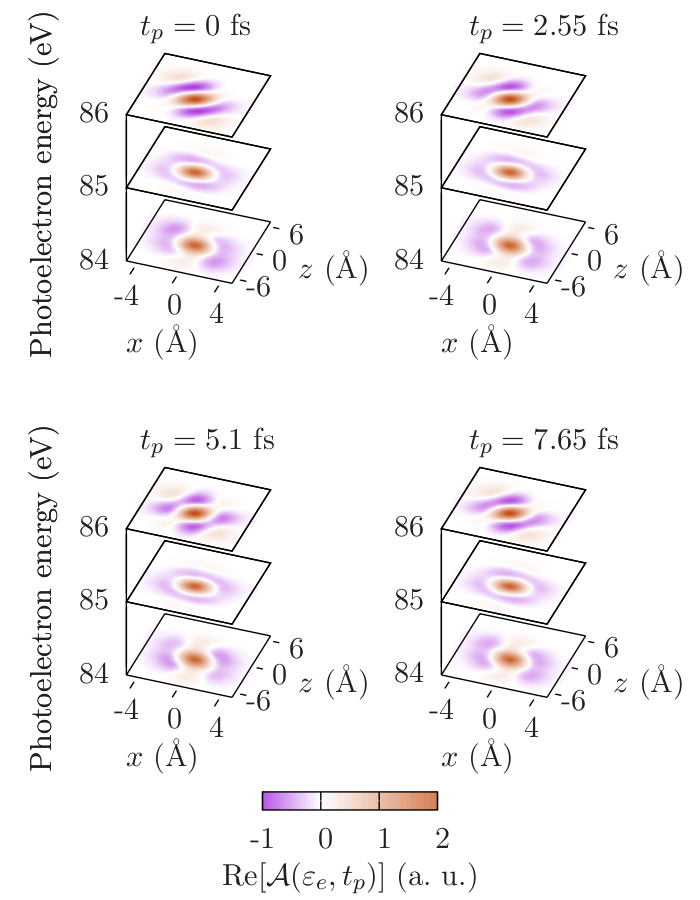}}\\
\subfloat[]{\includegraphics[width = 0.45\textwidth]{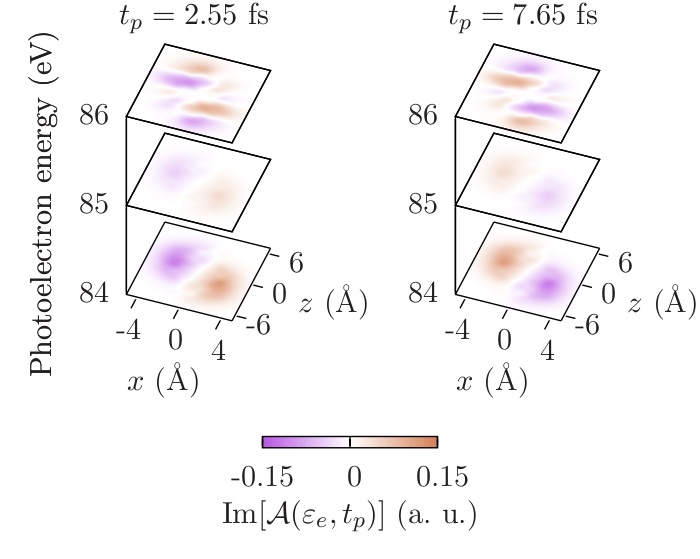}}
\caption{The real (a) and imaginary (b) parts of $\mathcal A(\text{84 eV},t_p)$, $\mathcal A(\text{85 eV},t_p)$ and $\mathcal A(\text{86 eV},t_p)$ at different times $t_p$.}
\label{Fig_orb-orb}
\end{figure}

Since the autocorrelation function of a time-dependent Dyson orbital, $A[\phi_F^D(t_p)](\mathbf r)$, has nonzero real and imaginary parts, the Fourier transform $\mathcal F(\mathbf r,\varepsilon_e,t_p)$ has also nonzero real and imaginary parts. The real and imaginary parts of the Fourier transforms $\mathcal F(\mathbf r,\varepsilon_e,t_p)$ of the angle-resolved spectra in Fig.~\ref{Fig_arspectra} at photoelectron energies 84 eV, 85 eV and 86 eV for $\mathbf r = (x,0,z)$ are shown in Figs.~\ref{Fig_F2D}(a) and \ref{Fig_F2D}(b), respectively. The imaginary parts are zero at $t_p=0$ fs and $t_p=5.1$ fs, and, thus, are shown only at $t_p=2.55$ and $t_p=7.65$ fs in Fig.~\ref{Fig_F2D}(b). At $t_p=2.55$ fs and $t_p=7.65$ fs, when the time-dependent electron density coincides, the real parts are identical and the imaginary parts have opposite sign. 


According to Eq.~(\ref{F_2D}), the Fourier transform of the time- and angle-resolved photoelectron probability at a fixed photoelectron energy $\varepsilon_e$ is determined by a linear combination of autocorrelation functions of Dyson orbitals $\phi_{F}^D(\mathbf r,t_p)$ with the coefficients given by the exponential factor in Eq.~(\ref{F_2D}). Only Dyson orbitals with corresponding final-state energies $E_F^{\text{ion}2+}$ spanning $\omega_\i+\la E\ra-\varepsilon_e$ within the bandwidth of the probe pulse contribute to the Fourier transform. These linear combinations at 84 eV, 85 eV and 86 eV are
\begin{align}
\mathcal A(\mathbf r,\text{84 eV},t_p) = & 1.2 A(\phi_{6}^D)+0.7A(\phi_7^D) +3A(\phi_{9}^D)\nonumber \\
&+0.6 A(\phi_{10}^D)\label{A84}\\
\mathcal A(\mathbf r,\text{85 eV},t_p) = & 0.2 A(\phi_1^D)+0.6A(\phi_{3}^D)+0.7A(\phi_4^D)\nonumber\\
&+3A(\phi_{6}^D)+0.9A(\phi_7^D) +0.9A(\phi_{9}^D)\label{A85}\\
\mathcal A(\mathbf r,\text{86 eV},t_p) = & 0.9 A(\phi_1^D)+2.7A(\phi_{3}^D)+0.9A(\phi_4^D)\nonumber\\
&+1.5A(\phi_{6}^D)+0.2A(\phi_7^D)\label{A86},
\end{align}
where  $\phi_{1-10}^D=\phi_{1-10}^D(\mathbf r, t_p)$ are the time-dependent Dyson orbitals corresponding to the states 1-10 in Table \ref{table_FinalStates}. Here, it is taken into account that $A(\phi_{2}^D) = 2A(\phi_{3}^D)$, $A(\phi_{5}^D) = 2A(\phi_{6}^D)$ and $A(\phi_{8}^D) = 2A(\phi_{9}^D)$. Thus, the Fourier transforms $\mathcal F(\mathbf r,\varepsilon_e,t_p)$ at $\varepsilon_e$ = 84 eV, 85 eV and 86 eV provide  linear combinations $\mathcal A(\mathbf r,\text{84 eV},t_p)$, $\mathcal A(\mathbf r,\text{85 eV},t_p)$ and $\mathcal A(\mathbf r,\text{86 eV},t_p)$ in the three-dimensional real space smeared out due to the convolution with $\operatorname{sinc}(q_0r)$.

According to Figs.~\ref{Fig_F2D}(a) and \ref{Fig_F2D}(b), the time-dependent part of $\mathcal F(\mathbf r,\varepsilon_e,t_p)$ is the largest at $\varepsilon_e = 86$ eV. Thus, the time-dependent photoelectron angular distribution at this energy most strongly correlates with the electron dynamics. $\mathcal F(\mathbf r,86\text{ eV},t_p)$ is determined by five autocorrelation functions of Dyson orbitals [see Eq.~\ref{A86}], among which the two autocorrelation functions $A(\phi_{3}^D)$ and $A(\phi_{4}^D)$ are time-dependent. Their real parts and, consequently, $\Re[\mathcal F(\mathbf r,\varepsilon_e,t_p)]$ are determined by the real parts of the density matrix elements, $\Re[C_IC_K^*]\cos((E_I-E_K)t_p)$ [see Eq.~(\ref{Probab_wp_Re_Im})] and correlate with related quantities such as the time-dependent electron density or instantaneous charge distributions. Their imaginary parts and, consequently, $\Im[\mathcal F(\mathbf r,\varepsilon_e,t_p)]$ are determined by the imaginary parts of the density matrix elements and correlate with the probability current density, which provides the electron current. Thus, by performing the Fourier analysis, one can distinguish between contributions to photoelectron angular distributions due to instantaneous charge distributions and due to the electron currents.

Note that the Fourier transform of a whole angle-resolved photoelectron spectrum would be determined by autocorrelation functions of Dyson orbitals corresponding to all possible final states [see Fig.~\ref{Fig_spectra}(b)], \textit{i.e.,} by a much larger number of autocorrelation functions  than at a fixed photoelectron energy. This would not only make the analysis more complicated, but would also strongly suppress the contrast, since many autocorrelation functions are time-independent. It is more advantageous to analyze photoelectron angular distributions at some fixed photoelectron energies, at which mainly time-dependent autocorrelation functions contribute.

The real and imaginary parts of the functions $\mathcal A(\text{84 eV},t_p)$, $\mathcal A(\text{85 eV},t_p)$ and $\mathcal A(\text{86 eV},t_p)$ for $\mathbf r = (x,0,z)$ are shown in Figs.~\ref{Fig_orb-orb}(a) and \ref{Fig_orb-orb}(b) in order to compare them with the Fourier transforms $\mathcal F(\mathbf r,\varepsilon_e,t_p)$ in Figs.~\ref{Fig_F2D}(a) and \ref{Fig_F2D}(b), respectively. Thereby, we analyze how much the convolution with the $\operatorname{sinc}(q_0r)$ function affects the signal from the indole molecule, the length of which is about 7 \r A. $\mathcal F(\mathbf r,\varepsilon_e,t_p)$ and $\mathcal A(\varepsilon_e,t_p)$ have a similar overall structure although the difference between these functions is quite visible. The effect of the limited spatial resolution can be well demonstrated with $\Im[\mathcal F(\mathbf r,\varepsilon_e,t_p)]$ and $\Im[\mathcal A(\varepsilon_e,t_p)]$ at $\varepsilon_e$ = 84 eV and 85 eV. Namely, due to the convolution with $\operatorname{sinc}(q_0r)$ function, the sign of $\Im[\mathcal F(\mathbf r,\varepsilon_e,t_p)]$ is inverted with respect to $\Im[\mathcal A(\varepsilon_e,t_p)]$. The regions where the sign of $\Im[\mathcal F(\mathbf r,\varepsilon_e,t_p)]$ flips have a width of approximately 1 \r A corresponding to the spatial resolution.

\section{Electron dynamics during the action of the probe pulse}
\label{Section_fast_indole}

In Sec.~\ref{SectionTDPP}, we discussed that the condition for the applicability of Eq.~(\ref{Photoelprob-wp}) is that the duration of the probe pulse is much shorter than the characteristic time scale of electron dynamics. In that case, the time-resolved photoelectron probability can be connected to the electronic wave-packet state at the time of ionization by the probe pulse. The scenario considered in the previous section satisfies that condition. Let us now consider the situation when the probe-pulse duration is comparable to the characteristic time scale of electron dynamics.

\begin{figure}
\centering
\includegraphics{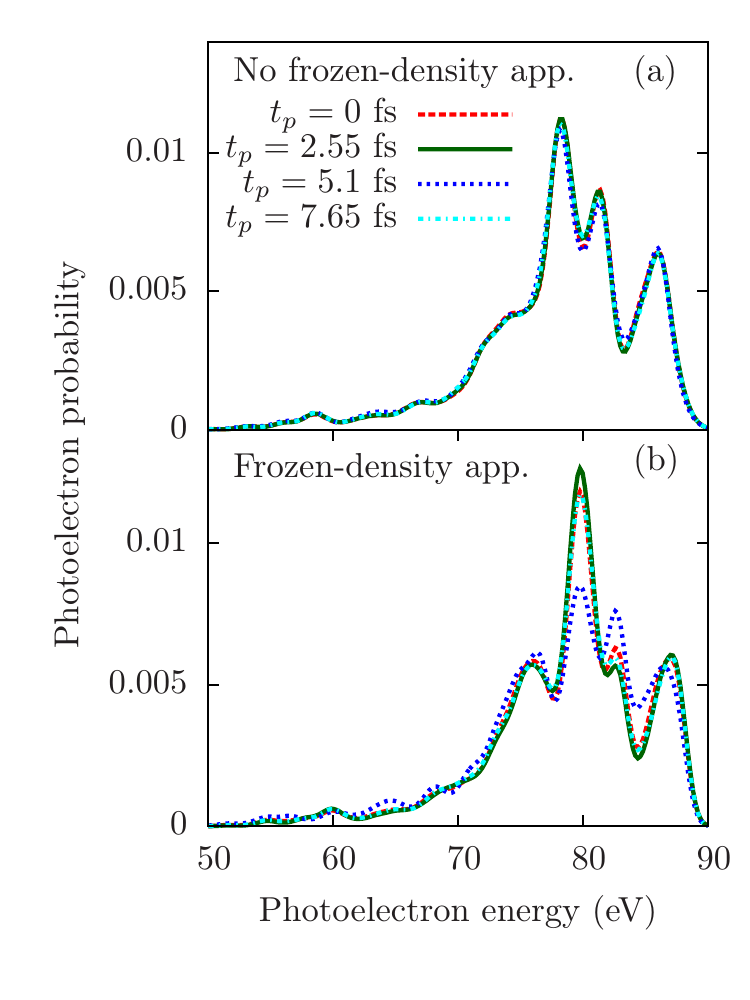}
\caption{Time-resolved photoelectron spectra in the case of three eigenstates involved in the electron dynamics of indole (a) computed with Eq.~(\ref{Photoel_prob_general}) and (b) computed within the frozen-density approximation according to Eq.~(\ref{Photoelprob-wp}).}
\label{Fig_fr_nofr}
\end{figure}

We assume that the pump pulse launches a coherent electronic wave packet by creating an electron hole in a superposition of HOMO, HOMO-1 and HOMO-2 of indole; the probe pulse has the same parameters as in the previous section. The characteristic time scale of electron dynamics is now determined by the largest energy splitting among the eigenstates of the wave packet, which is the energy difference between the HOMO-2 and HOMO orbitals of 2.7 eV within the Koopmans' theorem. Thus, the shortest beating period of the electron density is now 1.5 fs, which is still larger than the probe pulse duration of 1 fs. 

Figure~\ref{Fig_fr_nofr}(a) shows angle-averaged photoelectron spectra at four probe-pulse arrival times computed according to Eq.~(\ref{Photoel_prob_general}), taking into account the evolution of the electronic system during the action of the probe pulse. Comparing them to the photoelectron spectra computed at the same probe-pulse arrival times within the frozen-density approximation [Eq.~(\ref{Photoelprob-wp})], shown in Fig.~\ref{Fig_fr_nofr}(b), it is clear that the frozen-density approximation breaks down and provides a completely different result compared to that obtained with a correct treatment of the electron dynamics during the probe pulse. Angle-resolved photoelectron spectra calculated with and without the frozen-density approximation (not shown here) have even a completely different structure. Thus, the connection of the photoelectron probability to the electronic wave packet state at the time of the probe-pulse arrival is lost. 

This example demonstrates that if the probe-pulse duration is comparable with the shortest beating period of the electron density of a given system, the generated photoelectron spectra are not connected to its instantaneous electronic state. However, a probe-pulse duration being ten times shorter than the oscillation period of the electron density is enough to satisfy the frozen-density approximation, as we have verified using the example considered in Sec.~\ref{Section_slow_indole}.

\section{Conclusions}

We have described how time-dependent photoelectron spectroscopy can be employed for probing electron dynamics in molecules. We found that, quite counter-intuitively, electron dynamics cannot be followed by means of time-dependent chemical shifts, since time-dependent photoelectron spectra consist of a series of photoelectron peaks centered on time-independent positions. The amplitude of the photoelectron peaks is indeed time-dependent, but the time variation of their angle-averaged amplitudes is quite small. Therefore, angle-averaged photoelectron spectra hardly change in time. Angle-resolved photoelectron spectra, however, do show a clear dependence on electron dynamics and their temporal changes are considerable.

Looking into time-resolved photoelectron angular distributions at fixed photoelectron energies has several advantages. First, it facilitates their analysis, since a much smaller number of Dyson orbitals has to be considered than in the case of energy-unresolved distributions. Second, concentrating on photoelectron angular distributions at high photoelectron energies allows for a high spatial resolution. In the case of the indole molecule, it additionally simplified the study, since the orthogonalization correction to the plane-wave approximation was suppressed at high photoelectron energies. 

Angle-resolved spectra do not follow the instantaneous electron density, instead, there are two contributions to the angle-resolved spectra. The first contribution is determined by the real part of electron density matrix elements, and correlates with the electron density and related quantities such as charge distributions at the time of the probe-pulse arrival. The second contribution is determined by the imaginary part of electron density matrix elements and correlates with the instantaneous electron current. We have shown how to distinguish the two contributions to the angle-resolved spectra using Fourier analysis.

The connection of the time-resolved photoelectron spectra to the time-dependent quantities described above holds only if the probe pulse duration is much shorter than the shortest beating period of the electron density. Otherwise, this connection breaks down and electron dynamics during the probe pulse duration contribute a time-independent background to the spectra decreasing the signal contrast. Finally, applying the configuration-interaction method for the calculation of the dicationic states of indole, we have demonstrated that electron correlations of the remaining ion after the ionization by the probe pulse are crucial for giving rise to a time dependence of the photoelectron spectra.

Alternatively to ultrafast resonant x-ray scattering \cite{PopovaGorelovaPRB15_1,PopovaGorelovaPRB15_2}, one can use time- and angle-resolved photoelectron spectroscopy to obtain real-space information on instantaneous charge distributions and electron currents in molecules. Recent progress in attosecond science makes an experimental implementation of this technique feasible in the near future \cite{ChiniNature14}.

\section*{Acknowledgments}
We thank Giovanni Cirmi, Oliver M\"ucke, and Franz X.~K\"artner for stimulating discussions.	

\begin{appendix}

\section{Derivation of the time- and angle-resolved photoelectron probability}
\label{AppendixTDPP}

In this section, we derive the time- and angle-resolved photoelectron probability for single-photon ionization of an electronic system with Hamiltonian $\hat H_{\text{m}}$ with eigenstates $|\Phi_I\ra$ and eigenenergies $E_{I}$ by an ultrashort probe pulse arriving at time $t_p$. We assume that electronic system is in a coherent superposition of electronic states $\Psi(t)=\sum_IC_Ie^{-iE_I(t-t_0)}|\Phi^{N_{\el}}_I\ra$ launched by a pump pulse at time $t_0$, where $C_I$ are time-independent coefficients and $N_{\el}$ is the number of electrons in the system after the pump pulse and before the probe pulse (atomic units are used for this and the following expressions). The total Hamiltonian of the electronic system plus light is \cite{Loudon}
\begin{equation}
\hat H = \hat H_{\text{m}}+\sum_{\mathbf{k},s}\omega_{\k,s}\hat a_{\k,s}^\dagger\hat a_{\k,s}+\hat H_{\text{int}},
\end{equation}
where $\hat a_{\k,s}^\dagger$ and $\hat a_{\k,s}$ are creation and annihilation operators of a photon in the $\k$, $s$ mode with energy $\omega_{\k}=|\k|c$,  $c$ is the speed of light. 
\begin{equation}
\hat H_{\text{int}}=
\frac1c\int d^3r\hat \psi^\dagger(\mathbf r)\lf(\hat{\mathbf A}(\mathbf r)\cdot\mathbf p\rt)\hat \psi(\mathbf r),\label{H_int_App}
\end{equation}
is the minimal coupling interaction Hamiltonian in the Coulomb gauge, where $\hat{\mathbf A}$ is the vector potential operator of the electromagnetic field, $\mathbf p$ is the canonical momentum of an electron, $\hat \psi^\dagger$ and $\hat \psi$ are electron creation and annihilation field operators. 

The probability of observing an emitted electron with momentum $\mathbf q$ within the density matrix formalism \cite{Mandel} is given by
\begin{equation}
P=\lim_{t_f\to+t_f}\Tr\lf [\hat O_{\mathbf q}\hat\rho_f(t_f)\rt]\label{Probab_App},
\end{equation}
where $\hat\rho_f(t_f)$ is the total density matrix of the electron system and the electromagnetic field at time $t_f$ after the action of the probe pulse.
The operator 
\begin{equation}
\hat O_{\mathbf q} =\sum_{\sigma}c^\dagger_{\mathbf q,\sigma}c_{\mathbf q,\sigma}\label{Oks_App}.
\end{equation}
describes the observation of a photoelectron with momentum $\mathbf q$ and spin $\sigma$, $\hat c_{\mathbf q,\sigma}$ ($\hat c_{\mathbf q,\sigma}^\dagger$) creates (annihilates) an electron with momentum $\mathbf q$ and spin $\sigma$. The total density matrix within the first-order time-dependent perturbation theory using $H_{\text{int}}$ as the perturbation is
\begin{align}
\hat \rho^{(1,1)} =& \sum_{\{n\},\{\widetilde n\}}\rho ^X_{\{n\},\{\widetilde n\}}|\Psi_{\{n\}}^{(1)},t_f\ra\la\Psi_{\{\widetilde n\}}^{(1)},t_f|\label{rho11},
\end{align}
\begin{widetext}
where $|\Psi_{\{n\}}^{(1)},t_f\ra $ is the first-order wave function at time $t_f$, which is an entangled state of the electronic and photonic states:
\begin{align}
&|\Psi_{\{n\}}^{(1)},t_f\ra=-i\int_{t_0}^{t_f}dt\,e^{i(\hat H_{\text{m}}+\hat H_r) (t-t_f)}\hat H_{\text{int}}\,e^{-i(\hat H_{\text{m}}+\hat H_r) (t-t_0)}\Big|\{n\}\Big\ra\Big|\sum_IC_I\Phi^{N_{\el}}_I\Big\ra\label{Psi1_App}.
\end{align}

Thus, the photoelectron probability according to Eq.~(\ref{Probab_App}) is
\begin{align}
P =&\sum_\sigma \sum_{\{n\},\{\widetilde n\},I,K}C_IC_K^*\rho ^X_{\{n\},\{\widetilde n\}}\sum_{\k_1,\s_1,\k_2,\s_2 }\sum_{\{n'\},F}\frac{2\pi}{V\sqrt{\omega_{\k_1} \omega_{\k_2}} }\int_{t_0}^{t_f} dt_1\int_{t_0}^{t_f} dt_2  \\
&\times\la \{n'\} |e^{i\hat H_r (t_1-t_f)}\hat a_{\k_1,\s_1} e^{-i\hat H_r (t_1-t_0)}|\{n\}\ra
\la\{\widetilde n\}|e^{i\hat H_r (t_2-t_0)}\hat a^\dagger_{\k_2,\s_2}  e^{-i\hat H_r (t_2-t_f)} \ra |\{n'\} \ra\nonumber \\
&\times \la \Phi^{N_{\el}}_F |e^{i\hat H_{\text{m}} (t_1-t_f)}\hat T_{\k_1,\s_1}e^{-i\hat H_{\text{m}} (t_1-t_0)}|\Phi^{N_{\el}}_I\ra
 \la\Phi^{N_{\el}}_K|e^{i\hat H_{\text{m}} (t_2-t_0)}\hat T_{\k_2,\s_2}^\dagger e^{-i\hat H_{\text{m}} (t_2-t_f)} |\Phi^{N_{\el}}_F\ra\nonumber,
\end{align}
where $\hat T_{\k,\s} =  \int d^3 r\psi^\dagger(\mathbf r)(\boldsymbol\epsilon_{\k,\s}\cdot\mathbf p) e^{i\k\cdot\mathbf r}\psi(\mathbf r)$, which reduces to $\int d^3 r\psi^\dagger(\mathbf r)(\boldsymbol\epsilon_{\k,\s}\cdot\mathbf p) \psi(\mathbf r)$ within the dipole approximation.

We evaluate the final states $|\Phi_F^{N_{\el}}\ra$ assuming that the residual ion and the emitted photoelectron do not interact after the photoionization. Thus, $|\Phi_F^{N_{\el}}\ra = \hat c_{e,\sigma}^\dagger|\Phi_F^{N_{\el}-1}\ra$, where $|\Phi_F^{N_{\el}-1}\ra$ is an electronic state of residual ion with energy $E_F^{N_{\el}-1}$. Then,
\begin{align}
P =&\sum_\sigma  \sum_{\{n\},\{\widetilde n\},I,K}\rho ^X_{\{n\},\{\widetilde n\}}C_IC_K^*\sum_{\k_1,\s_1,\k_2,\s_2 }\sum_{\{n'\},F}\frac{2\pi}{V\sqrt{\omega_{\k_1} \omega_{\k_2}} }\int_{t_0}^{t_f} dt_1\int_{t_0}^{t_f} dt_2  \\
&\times e^{-i\omega_{\k_1} (t_1-t_0)}e^{i\omega_{\k_2} (t_2-t_0)} \la \{n'\} |\hat a_{\k_1,\s_1}|\{n\}\ra
\la\{\widetilde n\}|\hat a^\dagger_{\k_2,\s_2}   |\{n'\} \ra\nonumber \\
&\times e^{iE_K(t_2-t_0)}e^{-iE_I(t_1-t_0)}e^{i(-E_F^{N_{\el}-1}-\varepsilon_{e} )(t_2-t_1)} \la \Phi^{N_{\el}}_F |\hat T_{\k_1,\s_1}|\Phi^{N_{\el}}_I\ra
 \la\Phi^{N_{\el}}_K|\hat T_{\k_2,\s_2}^\dagger  |\Phi^{N_{\el}}_F\ra\nonumber
\end{align}
where $\varepsilon_{e}$ is the energy of the photoelectron.

\begin{figure*}
\centering
\includegraphics[width = \textwidth]{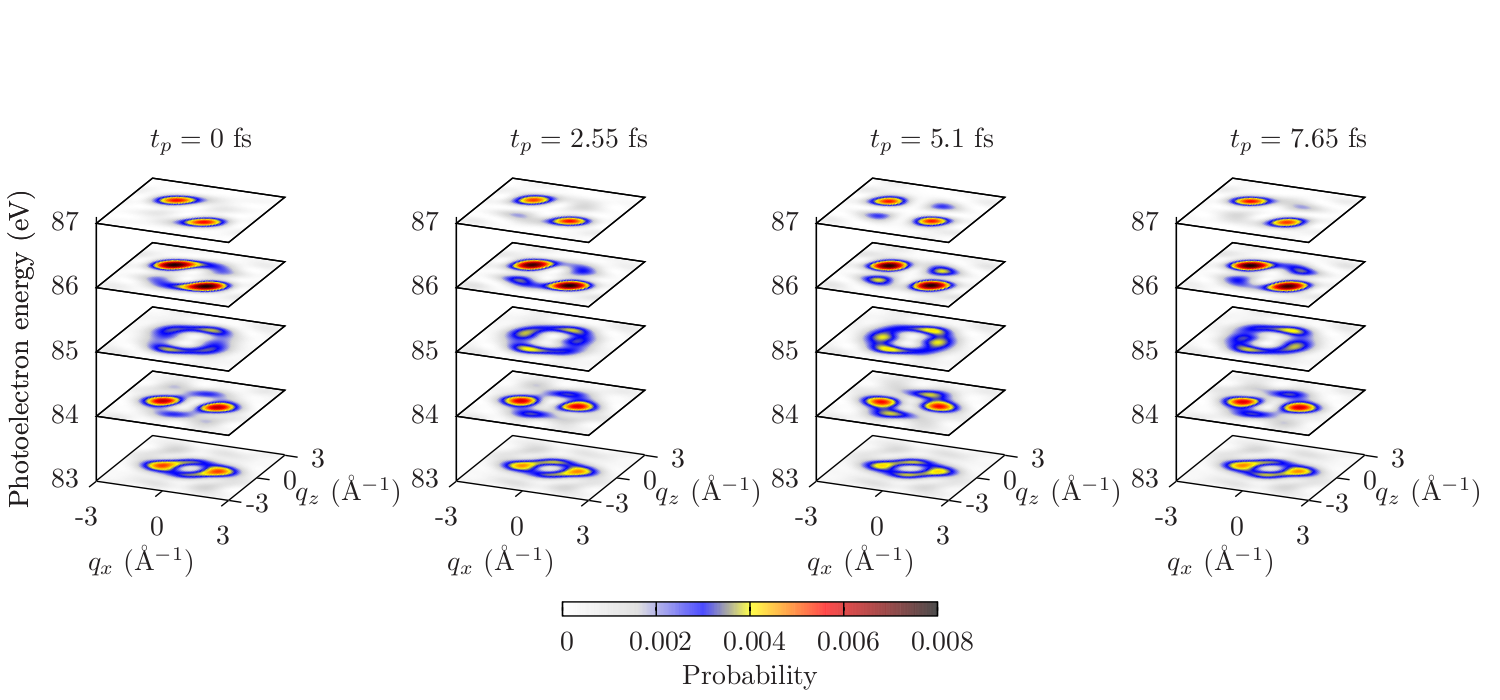}
\caption{Angle-resolved photoelectron spectra calculated taking into account the orthogonalization correction to the plane-wave approximation. Each $q_xq_z$ plane at the corresponding photoelectron energy $\varepsilon_e$ is a projection of a semisphere with radius $|\mathbf q| = \sqrt{2\varepsilon_e}$ at $q_y>0$, where the color of the $\mathbf q$ point on the semisphere corresponds to the probability of detecting an electron with momentum $\mathbf q$.}
\label{Fig_arpes_ortho}
\end{figure*}

In can be applied for a Gaussian-shaped probe pulse polarized along $\boldsymbol\epsilon_\i$ with the amplitude of the electric field $E(\mathbf r_0,t)=\sqrt{(8\pi/c) I_0(\mathbf r_0)}e^{-2\ln2\lf(\frac{t-t_p}{\tau_p}\rt)^2}$, where $\mathbf r_0$ is the position of the object, $t_p$ is the time of the measurement, $\tau_p$ is the pulse duration (FWHM of the pulse intensity) and $I_0(\mathbf r_0)=cE^2(\mathbf r_0,t=0)/(8\pi)$, that \cite{DixitPNAS12}
\begin{align}
&\frac{2\pi\omega_{\text{in}}}{V}\sum_{\mathbf k_1,\mathbf k_2, s_1,s_2}\sum_{\{n\},\{\widetilde n\}}\rho ^X_{\{n\},\{\widetilde n\}} \sum_{\{n'\}}\la\{n'\}|\hat a_{\k_1,s_1}|\{n\}\ra \la \{\widetilde n\}|  \hat a^\dagger_{\k_2,s_2}|\{ n' \}\ra e^{-i\omega_{\mathbf k_1}(t_1-t_0)}e^{i\omega_{\mathbf k_2}(t_2-t_0)}\\
&=\frac{2\pi}{c} I_0(\mathbf r_0)e^{-2\ln2\lf(\frac{t_1-t_p}{\tau_p}\rt)^2}e^{-2\ln2\lf(\frac{t_2-t_p}{\tau_p}\rt)^2}e^{i\omega_{\text{in}}(t_2-t_1)}\nonumber,
\end{align}
which results in 
\begin{align}
P =& \frac{2\pi I_0(\mathbf r_0)}{\omega_\i^2 c} \sum_{I,K}C_IC_K^*\sum_{F,\sigma} \la \Phi^{N_{\el}}_F |\hat T_{\k_\i,\s_\i}|\Phi^{N_{\el}}_I\ra
 \la\Phi^{N_{\el}}_K|\hat T_{\k_\i,\s_\i}^\dagger  |\Phi^{N_{\el}}_F \ra\\
&\times\Biggl(\int_{t_0}^{t_f} dt_1 \, e^{-2\ln2\lf(\frac{t_1-t_p}{\tau_p}\rt)^2}e^{-i(E_I-E_F^{N_{\el}-1}-\varepsilon_{e} +\omega_\i)(t_1-t_0)} \Biggr)\nonumber\\
&\times\Biggl(\int_{t_0}^{t_f} dt_2 \, e^{-2\ln2\lf(\frac{t_2-t_p}{\tau_p}\rt)^2}e^{i(E_K-E_F^{N_{\el}-1}-\varepsilon_{e} +\omega_\i)(t_2-t_0)} \Biggr)\nonumber.
\end{align}
Performing the integration and applying that
\begin{align}
&\la\Phi^{N_{\el}}_F|\hat T_{\k_\i,\s_\i}  |\Phi^{N_{\el}}_I \ra =\int d^3r  \phi_e^\dagger(\mathbf q,\mathbf r)(\boldsymbol\epsilon_\i\cdot\mathbf p) \phi_{F,I}^D(\mathbf r)\\
&\phi_{F,I}^D(\mathbf r) =\sum_\alpha\la \Phi_F^{N_{\el}-1}|\hat c_{\alpha}|\Phi_I^{N_{\el}}\ra\phi_{\alpha}(\mathbf r),
\end{align}
where $ \phi_e(\mathbf q,\mathbf r)$ is the photoelectron wave function, we obtain
\begin{align}
&P = \frac{\pi^2 \tau_p^2 I_0}{\ln2\omega_\i^2 c}\sum_{F,\sigma} \Bigg|\sum_{I}e^{-(\omega_\i-E_F^{N_{\el}-1} +E_{I}-\varepsilon_{e})^2\tau_p^2/(4\ln2)}\\
&\times\int d^3 r\phi_e^\dagger(\mathbf r, \mathbf q)\,(\boldsymbol\epsilon_\i\cdot\mathbf p)\la\Phi_{F}^{N_{\text{\el}}-1}|\hat\psi(\mathbf r)|C_Ie^{-iE_I(t_p-t_0)}\Phi^{N_{\el}}_I\ra\Bigr|^2.\label{Photoel_prob_general_App}
\end{align}

\end{widetext}
\section{Orthogonalization correction to the plane-wave approximation}
\label{Appendix_Ortho}

Figure \ref{Fig_arpes_ortho} shows the time- and angle-resolved photoelectron probability calculated with the photoelectron wave function
\begin{align}
|\phi_e\ra = |q_e\ra-\sum_{\alpha}|\phi_\alpha\ra\la\phi_\alpha|q_e\ra,
\end{align}
which is a plane wave $|q_e\ra=e^{i\mathbf q\cdot\mathbf r}/\sqrt{(2\pi)^3}$ orthogonalized to the molecular orbitals of the cation $\phi_\alpha(\mathbf r)$ \cite{RabalaisJChPh74}. The probe pulse parameters are the same as assumed for the calculation of the time- and angle-resolved photoelectron probability in Fig.~3 in the main text. The spectra are almost identical to the spectra in Fig.~3. A small discrepancy appears at large angles between photoelectron momentum vector $\mathbf q$ and $\boldsymbol\epsilon_\i$ in accordance with Refs.~\cite{PuschnigScience09, PuschnigChapter13, PuschnigJESRP15}.

\end{appendix}

\end{document}